\documentstyle[twocolumn,aps,prb]{revtex}

\def\tp{${\cal T}^{P}$}
\def\ts2f{${\cal T}^{*(2F)}$}
\def\tsa4{${\cal T}^{*(A_4)}$}\def\tfo{${{\mbox{$\bigcirc$}}\!\!\!\!\!\!\:{\mbox{2}}\,}$} 
\def\dfo{${{\mbox{$\bigcirc$}}\!\!\!\!\!\!\:{\mbox{2}}\,}$} 
\def\ffo{${{\mbox{$\bigcirc$}}\!\!\!\!\!\!\:{\mbox{5}}\,}$} 
\def\ep{${{\mbox{I}}\!{\mbox{{\bf E}$_\parallel$ } }}$}
\def\es{${{\mbox{I}}\!{\mbox{{\bf E}$_\perp $} }}$}

\begin{document}
\draft

\title{Surface structure of i--Al$_{68}$Pd$_{23}$Mn$_{9}$:
       An analysis based on the ${\cal T}^{*(2F)}$ tiling 
       decorated by Bergman polytopes}

\author{G.\ Kasner\\
Inst.\ f\"ur\ Theor.\ Phys.\ Uni--Magdeburg\\ 
PSF\ 4120,\ D--39016\ Magdeburg,\ Germany\\
Z.\ Papadopolos and P.\ Kramer\\ 
 Inst.\ f\"ur\ Theor.\ Phys.\ Uni-T{\"u}bingen,\\
Auf\ der\ Morgenstelle\ 14,\ D--72076\ T{\"u}bingen, Germany\\
D.\ E.\ B\"urgler \\
Inst.\ f\"ur\ Phys.\ Uni--Basel\\
Klingelbergstrasse 82, CH-4056 Basel, Switzerland}

\date{July 06 1998, revised version October 16 1998}
\maketitle

\widetext
\begin{abstract}
A Fibonacci--like terrace structure along a 5fold axis of 
i--Al$_{68}$Pd$_{23}$Mn$_{9}$  monograins has been observed by 
T.M. Schaub et al. with scanning tunnelling microscopy (STM). 
In the planes of the terraces they see patterns of dark 
pentagonal holes. These holes are well oriented both within 
and among terraces. In one of 11 planes Schaub et al. 
obtain the autocorrelation function of the hole pattern. 
We interpret these experimental findings in terms of the
Katz--Gratias--de~Boisseu--Elser model. Following the suggestion 
of Elser that the Bergman clusters are the dominant motive of 
this model, we decorate  the tiling \ts2f \ by the Bergman 
polytopes only. The tiling \ts2f \ allows us to use the powerful 
tools of the projection techniques.  The  Bergman polytopes 
can be easily replaced by the Mackay polytopes as the 
decoration objects. We derive a picture of ``geared'' layers 
of Bergman polytopes from the projection techniques as well 
as from a huge patch.  Under the assumption that no surface 
reconstruction takes place, this picture explains the 
Fibonacci--sequence of the step heights as well as the related 
structure in the terraces qualitatively and to certain
extent even quantitatively. Furthermore, this 
layer--picture requires that the polytopes are cut in order 
to allow for the observed step heights. We conclude that 
Bergman or Mackay clusters have to be considered as geometric 
building blocks of the i--AlPdMn structure rather than as 
energetically stable entities.
\end{abstract}

\pacs{PACS numbers: 61.44.Br, 68.35.Bs}

\narrowtext

\section{Introduction}
The surface of i--Al$_{68}$Pd$_{23}$Mn$_{9}$ perpendicular to 
5fold axes of an icosahedron has been explored in various 
papers, and terraces similar to netplanes in crystals have 
been observed~\cite{basel}.  
Schaub et al.~\cite{basel} obtained by scanning tunnelling 
microscopy (STM) atomic scale direct space information and 
low--energy electron diffraction (LEED) patterns of the 
sputtered and annealed quasicrystalline surface.  The dynamical 
LEED study of Gierer et al.~\cite{GiererT} of a similarly 
prepared surface confirmed the quasicrystalline structure 
and yielded additional structural information allowing an 
identification of the possible surface layers in terms of 
the bulk structure model by de Boissieu et al.~\cite{bois}.

A second STM study of in--situ cleaved surfaces by 
Ebert et al.~\cite{Urban} revealed terraces only  after 
annealing of the initially rather rough surface.

For the bulk structure of i--AlPdMn there exists a model due to 
de~Boissieu et al.~\cite{bois} which was 
generalised by Elser~\cite{elser} into the model that we refer to
as the Katz--Gratias--de~Boissieu--Elser model~\cite{elser,kg}. 
In order to obtain a geometric description, we consider
the Katz--Gratias--de~Boissieu--Elser model as a model for 
the {\em atomic positions}, independently of the particular 
chemical identity of the atoms (Al, Pd, or Mn) occupying these 
positions. \\
 
\vspace{7.2cm}
Therefore, all our conclusions will be valid for 
both, the de~Boissieu--Boudard model~\cite{bois} and the Elser 
model~\cite{elser} with the particular decoration of the 
atomic positions by Al, Pd, and Mn from the 
Katz--Gratias--de~Boissieu--Elser model.  The 
Katz--Gratias--de~Boissieu--Elser model consists of alternating 
Bergman and Mackay polytopes~\cite{elser}.  Elser proposed that 
the Bergman polytopes are not only {\em geometric clusters} but 
should also be considered as {\em energetically stable clusters}.
Our considerations are testing the conjecture.

 We interpret the Katz--Gratias--de~Boissieu--Elser model as the 
tiling ${\cal T}^{*(2F)}$~\cite{kram1} decorated by Bergman 
polytopes and some other, additional atomic positions 
(see~\cite{KPL297} and Section~\ref{geom}) forming Mackay 
polytopes.  As suggested by Elser~\cite{elser}, the dominant 
motives on this tiling model are dodecahedral Bergman clusters.
We adopt this suggestion (neglecting the additional atomic 
positions) and examine the layer stacking and the structure 
within planes perpendicular to a 5fold axis.  We compare 
the qualitative and quantitative predictions of the planar 
structure of the bulk model with the experimental findings 
at the surface~\cite{basel}.

The model analysis is made in terms of a patch of the 
tiling \ts2f~\cite{pap98} in 10$^{th}$ step of inflation 
decorated by Bergman polytopes.  This method allows to 
generate the relevant planar structure orthogonal to a 
5fold direction.  By the method of lifting we can relate the 
planar patch structure in \ep \ to the relevant 
triacontahedral window in \es \ with its coding substructure 
for the tiling \ts2f\ and find in it the coding for the planes.
An alternative approach to the terrace structure starting 
from the window side is given in Ref.~\cite{kpt}.

\section{Experiment}

The terrace--structure of the i--Al$_{68}$Pd$_{23}$Mn$_{9}$ 
monograin has been observed by STM~\cite{basel}. The terraces 
orthogonal to the 5fold axis are placed on Fibonacci 
distances with the short (low) interval L=4.22$\pm$0.26{\AA} 
and long (high) interval H=6.78$\pm$0.24{\AA}.
The planes (as terraces) occur in the sequence
        H H L H H L H L H H, see Figure~1.
	
In each plane there are dark pentagons (pentagonal holes)
oriented parallel to each other, both in a terrace and 
among the terraces~\cite{basel}.
In the terraces there are also white 5--stars with 5 dark 
pentagons between the star--arms, see Figure~2. 
In each 
plane one can draw lines through the structure such that 
they are  on Fibonacci distances, short (narrow) 
N=7.17$\pm$0.08{\AA} and long (wide) W=11.6$\pm$0.13{\AA}. 
The short interval N equals the height of the pentagonal holes.

The Patterson distribution function of the pentagonal holes
in the biggest  terrace No.~8 has been 
determined, see Figure~3.

\section{Geometric Model for the Atomic Positions}
\label{geom}

In order to see if the 
Katz--Gratias--de~Boissieu--Elser model~\cite{elser,kg}
of i--AlPdMn~\cite{bois} can 
explain the terrace structure, we consider this atomic model
(in the formulation by Elser~\cite{elser}), interpreted  
in terms of the canonical tiling \ts2f~\cite{kram1}, 
see Figure~4. 

The tiling \ts2f \  is related to the primitive tiling
\tp~\cite{kram1}  by converting only the tetrahedra 
$G^*$ ($G^*_{\parallel}$) and $F^*$ ($F^*_{\parallel}$) 
into acute and obtuse rhombohedra, respectively. 
The vertices of the tiling \ts2f \ then coincide with the
{\em even vertices} (even index sum) of the primitive tiling.

In
the decoration of the tiling \ts2f, Bergman polytopes (clusters) 
are centred~\cite{KPL297} at the {\em odd vertices} (odd 
index sum) of the primitive tiling, which are {\em not} 
vertices of the tiling \ts2f. It turns out that at least 
some pentagonal faces of each Bergman polytope (essentially a
dodecahedron) appear in the tiling \ts2f \ inscribed in 
the faces 
$\Sigma^*_2$ ($\Sigma^*_{2\parallel}$) and 
$\Sigma^*_3$ ($\Sigma^*_{3\parallel}$) of the tetrahedra, 
as shown in Figure~5 on an example of  $G^*$  and $F^*$  
tetrahedra. $\Sigma^*_2$  and $\Sigma^*_3$ are equilateral 
golden triangles {\em orthogonal} to the 5fold symmetry 
axes of an icosahedron. They are described in 
Section~\ref{geom1} 
together with other geometric properties of the tiling \ts2f.

\subsection{The planar structure of the tiling \ts2f}
\label{geom1}

We start with a description of the window in \es \ for the
tiling \ts2f \ and its coding content.
The window for the tiling is a triacontahedron. 
Our aim is to look for a possible coding of the planes
orthogonal to a 5fold direction of an icosahedron, 
that appear as a Fibonacci sequence on mutual distances as
observed in the experiment~\cite{basel}. These planes should 
contain the quasilattice points of \ts2f.
With respect to a fixed 5fold axis, we slice the 
triacontahedron into ten perpendicular zones
of five types (1, 2, ... 5) as shown in Figure~6.
The thickness of unions of these 
zones is:
  
\noindent
  $1=+1\cup -1 =\pm 1 \equiv x 
=\left( \frac{2}{\tau+2}\right)$ \ffo \\
  $\pm 1 \cup\pm 2 =  1 \cup 2 = \tau x$\\
  $\pm 1 \cup\pm 2 \cup\pm 3 = 1 \cup 2 \cup 3 = \tau^2 x$\\
  $\pm 1 \cup\pm 2 \cup\pm 3 \cup\pm 4 
    = 1 \cup 2 \cup 3 \cup 4 = \tau^3 x$ .\\ 
The symbol \ffo \ is the standard distance $\frac{1}{\sqrt2}$ 
along a direction parallel to a 5fold symmetry axis of an 
icosahedron, $\tau=\frac{1+\sqrt5}{2}$. The symbol \dfo \ is 
the standard distance $\sqrt\frac{2}{\tau+2}$ along a 
direction parallel to a 2fold symmetry axis of an icosahedron; 
the relation in length to the standard \ffo \ is : 
\dfo \ $=\frac{2}{\sqrt{\tau+2}}$ \ffo. 

Considering the thickness of zones and their combinations we 
look for
windows that would be necessary to code Fibonacci sequences of
planes perpendicular to the 5fold axis.
The decagonal middle zone of type 1 (decagonal prism) is the 
window~\cite{kram1} for the 
canonical planar tiling \tsa4 by two golden 
triangles~\cite{baake}. This fact has been denoted as
the dissectability of the \ts2f \ into  planar 
subtilings \tsa4~\cite{kram1}.
The thickness $x$ of the zone 1 (the decagonal prism) allows,
as a necessary condition, for coding a Fibonacci sequence of
planes with mutual distances 
in \ep \ $s \equiv s_{\parallel}= 
\tau^3 \left(\frac{2}{\tau+2} \right)$\ffo \ and 
$l \equiv l_{\parallel} = \tau s$ 

\vspace{0.3cm}
\noindent
$x=|s_{\perp}|+|l_{\perp}|$,

\vspace{0.3cm}
\noindent
where $x$ is to be understood as a window~\cite{pap98}.
The thicknesses for the unions of the zones 
$1 \cup 2 $, $1 \cup 2 \cup 3 $ and $1 \cup 2 \cup 3 \cup 4$
scale this window by up to three
powers of $\tau$. Consequently they may code the Fibonacci
sequences of planes in \ep \ three times inflated, 
respectively. The final sequence consists of  mutual distances 
$s = \left(\frac{2}{\tau+2} \right)$\ffo \ and $l = \tau s$.\\  

The significance of the zones in the tiling
is related to their content of windows for geometric 
objects other than quasilattice points: the 1D edges of the 
tiling have as windows the perpendicular projections
of dual 5--boundaries, and the faces of the tiling 
the perpendicular projections of dual 4--boundaries. 
We consider the edges and the faces of the tiling \ts2f \ in
the planes perpendicular to a fixed direction of a 5fold 
axes. They are coded by the corresponding dual boundaries
related to the same 5fold axes in \es.
In these planes there can appear only  two of four kinds of 
faces of the tiling, the golden triangles 
$\Sigma^*_{2 \parallel}$ and $\Sigma^*_{3 \parallel}$. 
They are equilateral triangles with one edge $\tau$\dfo \  and 
two edges \dfo \ , and one 
edge \dfo \ and two edges $\tau$\dfo \ , 
respectively. They are coded by   the 4--boundaries 
projected to \es \ , 
$\Sigma_{2 \perp}$ and $\Sigma_{3 \perp}$ 
respectively~\cite{kram1}.
Similarly the short edge \dfo \ $=\Omega_{1\parallel}$ and 
the long edge $\tau$\dfo \ $=\Omega_{2\parallel}$
of these two triangles have the windows 
$\Omega_{1\perp}$ and $ \Omega_{2\perp}$  
respectively~\cite{kram1}.
All the points of these projected dual boundaries, 
related to the particular 5fold direction, are located 
within the triacontahedron and moreover 
inside the union $1\cup 2 \cup 3 \cup 4$ and not in the
zone $5$. From their intersections with the first 
decagonal prism we 
get  the full coding for the  triangle pattern \tsa4\ in 
\ep. For the zones $2$,  $3$ and $4$ outside the
 decagonal prism we 
expect in \ep\ a gradual reduction in the
density of triangle faces and edges, due to the  disappearance 
of their coding in the zones. We also expect qualitative
 differences
in the pattern in the planes of different types. Finally in the 
zone $5$ we expect no such triangles and edges. These reductions 
should go along with a reduction of the density of quasilattice 
points for the tiling \ts2f.

Here we do not prove the sufficient condition for the existence
of  Fibonacci sequences of  planes. In Ref.~\cite{kpt} we
 identify 
the vectors which generate this Fibonacci sequence. Instead
 we check 
our expectations based on the necessary conditions from the 
side of 
the window on the finite patch in \ep \ obtained by the
 inflation 
procedure for the tiling \ts2f~\cite{pap98}. We inspect a 
10--step 
inflation patch. We cut the patch  of \ts2f \ in \ep \ by planes 
orthogonal to the 5fold direction whenever
there is a quasilattice point. There appear 318  planes on three
mutual distances: 
$\tau^{-1}\left(\frac{2}{\tau+2} \right)$\ffo \ ,
$\left(\frac{2}{\tau+2} \right)$\ffo\ and
$\tau \left(\frac{2}{\tau+2} \right)$\ffo \ .
The patch construction will allow us to simulate in detail 
the planar structure and its sequence and to compare it 
with experimental findings~\cite{basel}.

At the same time we wish to keep track  of the 
coding window
structure described above. For this purpose we pass 
to \es \ by the unique procedure of lifting 
the quasilattice points of the patch
into the triacontahedral 
window. The points within each plane of  \ep \
perpendicular to a 5fold axis are lifted into an intersection 
of the triacontahedron
with a plane shifted along and 
perpendicular to the corresponding 5axis in \es.
From the shift (with respect to the center of all the 
points of the patch)
we assign to each plane one of the ten zones 
$\pm1$, $\pm 2$, $\pm 3$, $\pm4$, $\pm5$ of the triacontahedron 
introduced before. In the generated patch there 
appear 5 families of  
planes in \ep, corresponding to the ten zones of 5 types 
in \es: 
The decagonal prism yields the densest planes with the golden 
triangle tiling. The zones 2, 3  and 4 give planes still 
containing golden triangles (and edges) of the triangle tiling. 
In contrast, the planes in \ep \  of the 
type 5 in \es \ contain only points of the quasilattice but 
neither edges nor faces (see Figure~7). 

In the patch, there are 234 planes of the 
types $1 \cup 2 \cup 3 \cup 4$ and they do appear in a 
Fibonacci sequence with a short $s$ and a long $l$ spacing. As 
expected
$s\equiv s_{\parallel}=\left(\frac{2}{\tau+2} \right)$\ffo\ ,
$l\equiv l_{\parallel}=\tau \left(\frac{2}{\tau+2} \right)$\ffo;
$|s_{\perp}| + |l_{\perp}|=\tau^3x$, 
where  $ \tau^3x$
is the thickness  of the window for the Fibonacci spacing.
Also the types $1 \cup 2 \cup 3 $, $1 \cup 2 $ and $1$
appear in corresponding 
by $\tau$, $\tau^2$, and $\tau^3$, respectively, inflated 
Fibonacci sequences. Finally the planes of type 5 are 
not part of the Fibonacci sequence,
and lead to the three distances among the planes of {\em all} 
types, 
$\tau^{-1}\left(\frac{2}{\tau+2} \right)$\ffo \ ,
$\left(\frac{2}{\tau+2} \right)$\ffo\ and
$\tau \left(\frac{2}{\tau+2} \right)$\ffo \ . 
The planes of type 5 carry a low density of quasilattice
points.  

In the Katz--Gratias--de~Boissieu--Elser model
 the scale is $\tau$--times
bigger than in \ts2f~\cite{KPL297}.  The short edge has 
the length $\tau$\dfo \  and the long edge $\tau^2$\dfo \ .
For i--AlPdMn, the standard length is
\ffo \ $=4.56${\AA}~\cite{bois}. Inserting it into the model the
two spacings of the planes $1 \cup 2 \cup 3 \cup 4$ become 
$\tau s\equiv L=\tau \left(\frac{2}{\tau+2}\right)$\ffo \ $
=4.08${\AA} and
$\tau l\equiv H
=\tau^2 \left(\frac{2}{\tau+2}\right)$\ffo \ $ = 6.60$ {\AA}, 
in agreement with the measured step heights~\cite{basel}.

\subsection{ The layers of the Bergman polytopes 
related to the planes of the \ts2f \ tiling}
\label{geom2}

In the 10 times inflated patch we find 20 sequences of 11 
planes such that each sequence could correspond to the observed 
11 terraces in the experiment~\cite{basel} on the distances
H H L H H L H L H H, see Figure~1. Let us take one of these
sequences, the one from 182--197 and plot it along the
sequence determined by the experiment, Figure~8.

In the sequence 182--197 there are 5 planes of type 5 that
are not observed in the measurement. The biggest terrace
observed in the experiment, denoted by No.~8~\cite{basel}
appears to correspond to the plane No.~192 of 
type 3 in the sequence. In our 20 sequences,
on the position of the plane No.~8 there appears 16 times
the plane of type 3, coded in the zone $-3$, and 4 times 
of type 2, coded in the zone $-2$. 
For all 20 sequences the first plane
is of type 4, coded in the zone $+4$ by the interval
($min(z_{\perp})$, $max(z_{\perp})$)$\sim(0.681, 0.825)$.
The whole zone $+4$ is coded by the interval 
($\tau^3x/2$,$\tau^2x/2$)$\sim(0.828, 0.521)$.
The coding interval of the plane equivalent to terrace No.~8 is 
$(-0.434, -0.290)\subset$ ($-\tau^2x/2$, $-x/2$).

So far we considered the points, edges and faces of the tiling 
in the sequence of planes orthogonal to a 5fold direction. 
Now we turn to the decoration by the Bergman  clusters,
as suggested by Elser~\cite{elser}.
The decoration of the tiling \ts2f \ by the Bergman polytopes is
performed as stated in Section~\ref{geom1} and Ref.~\cite{KPL297}.
As a final result, related to the planes of 
type 1--5 orthogonal to 
a 5fold axes, there appear  \emph{layers} of Bergman polytopes.
The edge of the Bergman polytope (dodecahedron) 
is $\tau^{-1}$\dfo \ $=2.96${\AA} and consequently the height 
of the dodecahedron, and the layers, 
is $\tau^2 \left(\frac{2}{\tau+2}\right)$\ffo \ $ = 6.60${\AA}. 
It equals the high spacing of the Fibonacci planes of 
type 1--4, $H= 6.60${\AA}.

In Figure~9 all layers of Bergman polytopes 
with two opposite pentagonal faces orthogonal to a 5fold 
direction (z--axis) in a part of the 10--step inflation patch 
are 
presented. The part of the patch contains 
the planes No.~177--197, such that it includes the 11 planes from
Figure~8. The length of horizontal lines  in the rows 
(B1), (B2), (B3) represents the height of a Bergman dodecahedron,
and their horizontal positions give the positions with respect 
to 
quasilattice planes. Horizontal lines to the right of a 
quasilattice 
plane denote Bergman polytopes hanging below the plane, 
horizontal lines to the left are Bergman polytopes
standing on the plane. 

In particular, 
(B1) are the layers of 
Bergman polytopes such that they are in between the planes
of type $\pm1,\pm2,\pm3,\pm4$, hanging from one plane and
standing on another.
(B2) are the layers of Bergman polytopes hanging from some of 
the planes of type $\pm1,\pm2,\pm3,\pm4$  and eventually
standing on a plane of type 5.
(B3) are the layers of Bergman polytopes standing on some of
the planes of type $\pm1,\pm2,\pm3,\pm4$ and eventually
hanging on a plane of type 5.
Hence the latter cannot be
interpreted as situated below any of the planes 
of type $\pm1,\pm2,\pm3,\pm4$ . The densities of all the 
layers, $\rho(B)$ ($\rho_b(B)$, $\rho_a(B)$) are 
written in the Figure~9 under the 
horizontal lines representing 
Bergman layers, see also Table 1. The layers of Bergman
polytopes are ``geared'' to each other. 

If we wish to interpret the observed terraces as the planes of
the type 1--4 and consider the Bergman polytopes as 
the clusters~\cite{elser},
then we relate to each terrace (plane) the layer of the Bergman
polytopes \emph{below} the plane, i.e. the layer of hanging 
Bergman clusters.
These clusters touch with a pentagonal face a plane of 
quasilattice points  from below. If that happens, the atomic 
position at the midpoint
of the face is lowered with respect to the plane by $0.48$\AA,
occupied by Al in the $B_5$ position of the model~\cite{bois}. 
This face 
could appear as a dark hole in the STM experiment. The search 
for these pentagonal faces within the planes of type 1--4 is 
equivalent to the search for those  Bergman clusters which 
hang below these planes (from the layers B1 and B2). Knowing 
the coding interval in the window for the
terrace No.~8, $(-0.434, -0.290)\subset$ ($-\tau^2x/2$, $-x/2$),
one can, as shown in Ref. \cite{kpt}, compute the density of the
corresponding hanging Bergman polytope layer.

Using this approach we find the
density of the terrace No.~8 to be in the range 
of 5.72--8.62 $\cdot 10^{-3}$ hanging 
Bergman clusters/{\AA}$^{2}$.

As we already stated, the planes of type 1--4 are by their
mutual distances in agreement with the terraces observed by 
STM~\cite{basel}. The planes of type 5 are not observed as
terraces, probably due to the low densities of the quasilattice
points in the planes. How is the appearance of the terraces 
related to the Bergman layers? Planes (terraces) appear to be
correlated to  3 Bergman layers such that one layer is above
the plane, another below the plane and the third one is 
dissected by the plane. These planes 
are of type 1--4. For planes which appear correlated to only 2
Bergman layers such that w.r.t. the previous case either the
layer above or below the plane is missing, do not 
appear. These planes are of type 5.

%
%
\begin{table}
\begin{center}
\caption{ The densities of the quasilattice points $q$,
$\rho(q)$ of the tiling \ts2f\ in the planes No.~177--197
of the 10--times inflated patch. $\rho(q)$ is normalized w.r.t.
planes coded by the decagonal prism in
triacontahedron. The symbol $\eta$ is the normalized
$z$--coordinate in \es, $\eta=z/(\tau$ \ffo), the coordinate of
the  plane--coding in the window for \ts2f,
$\eta \in (-1,1) $. $\rho_b(B)$ and $\rho_a(B)$ are
respectively the densities of the Bergman--polytope layers
below and above the planes w.r.t. the direction of
the $z$--axis in \ep. $\rho(B)$ are normalized w.r.t. the
layers with the maximal density,
see  \cite{kpt}, where the meaning of the expected
sharp values  $0$ and $1$ in the brackets is explained.
The corresponding 11 terraces are situated
between the planes 182--197 as in Figure~8.}

\begin{tabular}[h]{cccccc}
Plane No. & Type & $\eta(q)$ & $\rho(q)$ & $\rho_b(B)$ & $\rho_a(B)$\\
\hline             
 $177$ & $-1$& $-0.050$ & $ 1.00$ & $0.95$ & $0.99$\\
$178$ & $5$& $0.845$ & $0.10$ & $0.62$ & $0.00 \; (0)$\\
$179$ & $-4$& $-0.603$ & $0.55$ & $0.05$ & $0.95$\\
$180$ & $3$& $0.292$ & $0.97$ & $1.00 \; (1)$ & $0.62$\\
$181$ & $-2$& $-0.261$ & $0.98$ & $0.66$ & $ 1.00 \; (1)$\\
$182$ & $4$& $0.633$ & $0.51$ & $0.92$ & $0.04$\\
$183$ & $-5$& $-0.814$ & $ 0.13$ & $0.00 \; (0)$ & $0.66$\\
$184$ & $1$& $0.081$ & $1.00$ & $1.00$ & $0.92$\\
$185$ & $5$& $0.975$ & $0.00$ & $0.37$ & $0.00 \; (0)$\\
$186$ & $-4$& $-0.472$ & $0.76$ & $0.25$ & $1.00$\\
$187$ & $3$& $0.422$ & $0.85$ & $1.00 \; (1) $ & $0.37$\\
$188$ & $-1$& $-0.131$ & $1.00$ & $0.86$ & $1.00$\\
$189$ & $5$& $0.764$ & $0.23$ & $0.75$ & $0.00 \; (0)$\\
$190$ & $-4$& $-0.683$ & $0.39$ & $0.01$ & $0.86$\\
$191$ & $2$& $0.211$ & $0.99$ & $1.00 \; (1)$ & $0.75$\\
$192$ & $-3$& $-0.342$ & $0.92$ & $0.51$ & $1.00 \; (1)$\\
$193$ & $4$& $0.553$ & $0.66$ & $0.97$ & $0.13$\\
$194$ & $-5$& $-0.894$ & $0.04$ & $0.00 \; (0) $ & $0.51$\\
$195$ & $1$& $0.000$ & $1.00$ & $0.98$ & $0.97$\\
$196$ & $5$& $0.894$ & $0.05$ & $0.53$ & $0.00 \; (0)$\\
$197$ & $-4$& $-0.553$ & $0.64$ & $0.10$ & $0.98$\\
\end{tabular}
\end{center}
\end{table}

\subsection{Interpretation of the pentagonal holes in the planes}
\label{geom3}

The observed dark pentagonal holes~\cite{basel} of the estimated 
height $7.17 \pm 0.08${\AA} are approximately $\tau$ times 
bigger than the pentagonal faces (face pentagons, see Figure~5) 
of the Bergman polytopes.  The height of the Bergman face 
within the plane is 4.56{\AA}.  The observed pentagonal holes 
are as big as the pentagons on a parallel cut through five 
vertices of the dodecahedron with identical orientation, see 
Figure~5.  Their height is 7.38{\AA}.  We call them top 
equatorial pentagons.  Such a pentagonal cut through a 
Bergman cluster would again have a midpoint, in this case 
lowered by $0.78${\AA}, occupied by Pd according to the 
model~\cite{bois}.  Such a pentagon could also appear as a 
hole.  The planes in the tiling and in the patch which contain 
these top equatorial pentagons are shifted with respect to the 
former planes by $(\frac{2}{\tau +2})$\ffo $ \ = 2.52${\AA}.  
Tentatively we propose this alternative interpretation of 
the pentagonal holes in the planes.  The identification of the 
pentagonal holes as top equatorial pentagons of Bergman 
polytopes is appealing because it readily explains the size of 
the pentagons.  However, there is a disagreement with the 
separation of the two topmost layers determined in 
a LEED--IV analysis by Gierer et al.~\cite{GiererT}. They find 
a separation of 0.38{\AA}, which is interpreted as a contracted 
bulk layer separation of 0.48{\AA}.  This value, in turn, would 
nicely fit the depth of the Bergman ``faces".  
Therefore, a clear--cut interpretation of the pentagons 
observed by STM is still lacking.

It is important to note that if we wish to relate the 
experimental data with the
 Katz--Gratias--de~Boissieu--Elser model, this 
implies in any case that the Bergman polytopes of height 
$\tau^2(\frac{2}{\tau + 2})$\ffo \ $=6.60${\AA}
are cut by the terrace structure with a minimal layer separation
of 4.22{\AA}, see Figure~9. 

The Mackay polytopes  of Katz--Gratias--de~Boissieu--Elser 
model~\cite{elser}  would also provide pentagonal holes. 
Their height would be 7.38{\AA}, but they are much deeper 
(2.52{\AA}), and Mn atoms  on $M_0$ positions should be in the 
center~\cite{bois}. 

The lines analyzed  in \cite{basel}  in a fixed plane 
with pentagons (see Figure~2) can be understood in the model 
as follows: 
Take another 5fold axis at an angle
$\alpha$ (see Figure~10), $\cos{\alpha} = \frac{\tau}{\tau+2}
=\frac{1}{\sqrt{5}}$ w.r.t. the fixed one (chosen as z--axes) 
and consider its set of planes of type $1\cup 2\cup 3$.
These planes will intersect the fixed plane in parallel lines 
in Fibonacci spacing with distances  N and W, where 
   $N = \frac{\sqrt{5}}{2} H = 7.38${\AA} and 
   $W = \frac{\sqrt{5}}{2} (L+H)=11.94${\AA}, 
$\sin{\alpha} = \frac{2}{\sqrt{5}}$, see Figure~10. 
These distances compare well with the experiment~\cite{basel}.    

From  Figure~8 we see that the terrace No.~8, on which 
the Patterson distribution function 
of the pentagonal holes (Figs.~2 and 3) was 
determined~\cite{basel}, corresponds to the plane No.~192
in the sequence of planes No.~182--197 of the 10--times 
inflated patch.
         
In  Figure~11 only those golden triangles of the plane 
No.~192 are presented, from which Bergman 
polytopes are hanging. They are hanging w.r.t. the 
positive direction of the z--axis of  Figs.~8 and 9. These 
Bergman polytopes 
are placed between the planes No.~192 and 194, below  192 and
above 194.

In Figure~12 we show the patterns of the top equatorial 
pentagons of  Bergman polytopes in the planes No.~188, 190, 
191, and 192.

These planes represent the terraces No.~5, 6, 7 and 8,
respectively.  The pentagons are oriented parallel to
each other, both in a terrace and among the terraces,
as observed in  Ref. \cite{basel}. Big fluctuations in the 
density of the Bergman polytopes in the layers is expected, 
see also Table 1. 

In order to compare our model to the experimentally
obtained results on the distribution of the dark
pentagonal holes in the STM measurement, we calculate
the Auto--Correlation Function (ACF) or Patterson 
function \newline

$A(\vec{r}) = \int\limits_{\cal
F}z_h(\vec{r'})z_h(\vec{r'}+\vec{r})d^2r'
{,}$\\

where $\vec{r}=(x,y)$ and $z_h(\vec{r})$ denotes the hole image.
The ACF for the distribution of the dark pentagons was 
calculated by digitizing plane No.~192 of the model in exactly 
the same manner as described in \cite{basel} by assigning the 
value 1 to those parts of the plane inside a pentagon and 0 
otherwise. The resolution was also chosen to coincide with 
the one used in \cite{basel}, namely 0.5 {\AA} per pixel. 
Numerically we obtained the ACF for plane No.~192 of size 
$764\times764${\AA$^2$}. The layer below the plane No.~192 in
 the 
patch contains 3835 hanging Bergman polytopes
and hence, the density of the pentagons in the plane Nr.~192 
is 6.58$\cdot 10^{-3}${\AA$^{-2}$}. 
With respect to noise in the STM images, local density
 fluctuations 
in small patches of a quasiperiodically decorated plane and
 freedom 
in the choice of the greyscale--level (which separates
 between black 
and white in the digitizing procedure of the STM--pattern) the 
estimated density on terrace No.~8 of
4.22$\cdot 10^{-3}${\AA$^{-2}$} can be considered to be
 in rather 
good agreement with that obtained from our model.
The minimal distance between the pentagons is the short 
edge of the tiling \ts2f.
In the Katz--Gratias--de~Boissieu--Elser model 
it equals $\tau$\tfo \ = $r_{calc}(I') \approx 7.8$ {\AA}.
The mean distance for the pentagons in the plane No.~192 is
 calculated 
to 12.33 {\AA}.

In Figure~13 the resulting ACF is shown for a range of the
displacement vectors of $\pm$ 100 {\AA} in $x$ and $y$ 
directions.
Labels on the first ten maxima are in correspondence with 
those of Figure~4 in Ref. \cite{basel} and Table~2.
The calculated peak positions fit well to those 
obtained from the hole pattern extracted from STM measurement,
see Figure~3 and Table~2. 

%
%
TABLE II. Radii of the Patterson correlation maxima in Figure~3
$r_{exp}$ and in Figure~13 $r_{calc}$.\\
\begin{tabular}[s]{c|cccccccccccc}
\hline\hline
 r[{\AA}] & I$'$ & I & II & III & IV & V & VI & VII & VIII & IX & X \\
\hline
$exp$ & & $\approx 12$ & $19.7$ & $31.7$ &
$36.9$ & $41.3$ & $49.4$ & $51.0$ & $60.5$ & $63.3$ & $68.1$\\
$calc$ & $7.8$ & $12.6$ & $20.3$ & $32.9$ &
$38.6$ & $43.7$ & $50.7$ & $53.2$ & $62.5$ & $65.7$ & $66.7$\\
\hline\hline
\end{tabular}

\section{Discussion}

In this paper we have used the projection--tools related to the 
tiling \ts2f. We have refined  the already known 
dissectability property~\cite{kram1,pap98} of the 
tiling \ts2f \  along the 5fold direction and have transfered 
this inherent property of the tiling  into the layer structure 
of Katz--Gratias--de~Boissieu--Elser model.

In addition, we have generated a huge patch of the tiling 
\ts2f \ by a highly non--trivial inflation 
procedure~\cite{pap98} to linear dimensions of about 750 {\AA}. 
It is large enough to reproduce all statistical predictions 
about densities from the projection method as well as to 
contain inflation symmetries (10$^{th}$ step of the inflation).

We do not consider the choice of the 
Katz--Gratias--de~Boissieu--Elser model as a significant 
restriction because most of the results derived above will 
also hold for models with approximately the same size of windows,
as for example the model from Janot~\cite{janot} that we have 
not yet considered. We have focused on the 
Katz--Gratias--de~Boissieu--Elser model because it is defined 
as a decoration of the tiling (primitive \tp, or \ts2f) 
and all powerful tools known from the projection methods are 
applicable.

\section{Conclusion}

We prove that the experimentally observed succession of the step 
heights L (low) and H (high) along the 5fold axis (z--direction),
which obeys the Fibonacci sequence, also exists in the patch of 
the geometric Katz--Gratias--de~Boissieu--Elser model. 
Additionally, we relate this sequence to another Fibonacci 
sequence of the distances N (narrow) and W (wide) between lines 
within the planes (x--y planes) of the terraces also found in 
experiments~\cite{basel}.

The estimated coding of the observed finite Fibonacci 
subsequence along the 5fold direction restricts the choice of 
planes in the model which have to be compared with the 
experimentally analysed terraces. Our model predicts big 
variations of the densities of quasilattice points 
(or equivalently pentagonal holes) among terraces which 
should be measurable in future high--quality STM images.

Our analysis shows that the Katz--Gratias--de~Boissieu--Elser 
model can be understood as being composed of ``geared'' layer 
of Bergman polytopes (Figure 9). We relate sequences of these
layers of Bergman polytopes  to the 
observed terraces and derive the patterns of pentagons within 
the terraces. We calculate the densities of Bergman clusters 
(Table 1) within the layers
based on the knowledge of the window for the Bergman 
clusters~\cite{kpt} as well as using our huge patch (Figure 9). 
The calculated pentagon distribution yields a reasonable mean 
density and good agreement of the Patterson data (Figure 13, 
Table 2).  The direct space patterns even reproduce some 
structures (white 5--stars) observed on the surface of the 
terraces.  Other features of the patterns (Figures 11 and 12) 
should be observable in future high-quality STM images.  Hence, 
we predict more detailed criteria to judge from STM data 
whether the Katz--Gratias--de~Boissieu--Elser model is 
realized or not.  

If we assume that the surface of i--AlPdMn is not reconstructed 
w.r.t. the bulk, a fact that follows from the work of Gierer et 
al.~\cite{GiererT}, then the Bergman polytopes represent a 
correct geometric decoration, but they may not be considered as 
energetically stable clusters.  This follows from the picture of
``geared'' layers of Bergman polytopes presented in Figure 9 
which requires that Bergman clusters are cut in order to allow 
for the observed step heights (H and L). Further, 
from the alternating 
decoration with Bergman and Mackay polytopes of the primitive
tiling~\cite{elser} one can also easily conclude that  the 
same model cannot be interpreted as the Mackay--cluster model 
either. Therefore, the Bergman and Mackay clusters have to be 
considered as geometric building blocks of the quasicrystalline 
structure rather than as energetically stable entities.  

\section*{Acknowledgements}

Financial support of the DFG is gratefully acknowledged 
by  Z.~Papadopolos and P.~Kramer. Z.~Papadopolos 
acknowledges support from the Centre de Physique Th\'{e}orique, 
CNRS, Marseille where she has done part of this work.
We would also like to thank the Geometry--Center at the 
University of Minnesota for making Geomview freely available, 
which proved to be a valuable tool throughout our work.

\section*{Figure captions}

%
%
\begin{figure}
\caption{The terrace structure of the i--Al$_{68}$Pd$_{23}$Mn$_{9}$
monograin~\cite{basel}}
\end{figure}

%
%
\begin{figure}
\caption{Atomic--scale structure within the terraces~\cite{basel}.
Data taken on terrace No.~8 of Figure~1. The white 5--star is marked
by an arrow.}
\end{figure}        

%
%
\begin{figure}
\caption{ Patterson distribution function of the pentagonal holes
on the terrace No.~8~\cite{basel}. The $x$ and $y$ axes extend
from -100 {\AA} to +100 {\AA}.}
\end{figure} 

%
%
\begin{figure}
\caption{ The window of the tiling \ts2f \, is a triacontahedron,
the six tetrahedra are the tiles~\cite{kram1}. The
symbols \ffo \ and \dfo \ are
the standard lengths defined in Section~2.~1}
\end{figure}  

%
%
\begin{figure}
\caption{The Bergman polytopes are typically hanging from the
golden triangles $\Sigma^*_2$ and $\Sigma^*_3$ as the faces
of the tiling \ts2f. Pentagonal faces of the Bergman
polytope (essentially a dodecahedron) and  top equatorial
pentagons, bigger by  a factor $\tau$, are marked.}
\end{figure}

%
%
\begin{figure}
\caption{Window (triacontahedron) sliced into ten perpendicular
zones orthogonal to a 5fold axis. The  point zero is at the
center of the triacontahedron. W.r.t. this  point zero, the
zones are denoted by $\pm 1, \pm2, \pm3, \pm4$ and $\pm5$.}
\end{figure} 

%
%
\begin{figure}
\caption{Examples in \ep of the planar sections orthogonal
to the 5fold axis of the tiling \ts2f, such that the planes
contain quasilattice points, the tiling vertices. Types 1--5
are coded in the window by the corresponding zones
$\pm 1,\pm2,...,\pm5$,
see Figure~6. The golden triangles $\Sigma^*_{2 \parallel}$
and $\Sigma^*_{3 \parallel}$  define the
structure of the planes of Type 1--4.}
\end{figure}          

%
%
\begin{figure}
\caption{In the 10--times inflated patch we find a sequence that
corresponds to the 11 terraces observed by STM. The plot
shows a histogram of Figure~1 (NODV = number of data values).
The numers and the types (zones) of the planes in the patch
and the number of the corresponding terrace in the STM image
(Figure~1) are indicated above the plot. Note that the planes
of types $\pm$5 are not observed in the experiment.}
\end{figure}    

%
%
\begin{figure}
\caption{All layers of Bergman polytopes orthogonal to a 5fold
direction (z--axis) in a part of the 10--step inflation patch
containing the planes 177--197.
The dotted intervals mark the relative
distances between the planes and the height of the Bergman
polytopes. The height of the Bergman
polytope equals the length $H$, where $H=6.60$\AA. }
\end{figure}    

%
%
\begin{figure}
\caption{Possible relation of the Fibonacci spacings of the
planes (terraces) of type 1--3  with the Fibonacci spacings
of the lines in the planes  based on the \ts2f\ tiling.}
\end{figure}    

%
%
\begin{figure}
\caption{ The representation of the terrace No.~8 by
the plane No.~192 of the 10--times inflated \ts2f\ patch.
Note the pentagonal faces or $\tau$ bigger top equatorial
pentagons
of Bergman polytopes, the lines on N and W distances (see
Figure~10) and the white star
of height 2W+N. Compare to  Figure~2.}
\end{figure}          

%
%
\begin{figure}
\caption{ The representation of the terraces No.~5, 6, 7 and 8 by the
planes No.~188, 190, 191 and 192 respectively. The content are
the golden triangles and the top equatorial pentagons of
Bergman polytopes.}
\end{figure}  

%
%
\begin{figure}
\caption{The Patterson distribution
function (the correlation maxima) of the pentagonal holes
in the plane Nr. 192. Compare to Figure~3. }
\end{figure}

\end{document}